\documentclass[twocolumn,showpacs,pre,floatfix,amsmath,amssymb,notitlepage]{revtex4-1}
\usepackage{epsfig}
\usepackage{subfigure}
\usepackage{amsmath}
\usepackage{color}
\usepackage{amssymb}
\usepackage{setspace}
\usepackage{graphicx}
\usepackage{dcolumn}
\usepackage{bm}
\usepackage{times}
\usepackage{enumerate}
\usepackage{footnote}
\usepackage{upgreek}
\input epsf

\graphicspath{{figures/}}

\begin{document}
%

\author{Linpeng Gu$^{1}$}
\author{Qingchen Yuan$^{1}$} 
\author{Qiang Zhao$^{2}$}
\author{Yafei Ji$^{1}$} 
\author{Ziyu Liu$^{2}$}
\author{Liang Fang$^{1}$} 
\author{Xuetao Gan$^{1*}$, Jianlin Zhao$^{1}$} 
\email{zhaoqiang@qxslab.cn; xuetaogan@nwpu.edu.cn} 
\affiliation{$^{1}$Key Laboratory of light-field manipulation and information acquisition, Ministry of Industry and Information Technology, and Shaanxi Key Laboratory of Optical Information Technology, School of Science, Northwestern Polytechnical University, Xi'an 710072, China}
\affiliation{$^{2}$Qian Xuesen laboratory of Space Technology, China Academy of Space Technology, Beijing 100094, China}

\date{\today}

\title{A topological photonic ring-resonator for on-chip channel filters}
%
%

\maketitle

\maketitle

{\bf

A topologically protected ring-resonator formed in valley photonic crystals is proposed and fabricated on a silicon slab. The unidirectional transmission and robustness against structure defects of its resonant modes are illustrated. Coupled with topological waveguides, the topological ring is functioned as notch and channel-drop filters. The work opens up a new avenue for developing advanced chip-integrated photonic circuits with attributes of topological photonics.
}\\

\section{Introduction}
%
%
%
%
Recently, topological photonics has attracted remarkable attention due to its unique properties~\cite{wang2009observation,shalaev2019robust,lu2014topological,khanikaev2017two,he2019silicon,gong2020topological}. Photonic analogs of topological quantum systems can be easily realized via gyro photonic crystal (PC) with an available promising platform~\cite{wang2008reflection}. A variety of structures have been designed to emulate the quantum Hall effect, quantum spin- and valley-Hall effects in electronics by introducing the concept of pseudo-spin degrees of freedom~\cite{haldane2008possible,raghu2008analogs,ma2016all,chen2017valley,dong2017valley,wu2015scheme,yang2018visualization}. Besides, according to the $bulk-boundary$ correspondence, the topologically protected edge waves existing at the interface between two photonic topological insulators exhibit excellent robustness, defect immunity, backscatter suppression and unidirectional transmission properties~\cite{shalaev2019robust,he2019silicon,khanikaev2013photonic,chen2014experimental,ma2015guiding}. Based on this, boundary topological channels have been constructed using two-dimensional topological photonic crystals (2D-TPC) and potentially employed in a number of promising scenarios, like high-efficiency non-reciprocal lasing with pure mode and low threshold~\cite{bahari2017nonreciprocal,smirnova2020room,shao2020high,zhang2020low}, topological photonic routing~\cite{he2019silicon,gong2020topological,liu2020z2,wang2020valley,tian2020dispersion}, topological all-optical logic gates~\cite{he2019topological}, robust delay lines~\cite{ma2016all,hafezi2011robust}, and so on. In addition, due to the topological protection characteristics, light can smoothly transport but sharply turn along the topological channel. Thus, a compact optical topological cavity with an arbitrary shape can be designed~\cite{gong2020topological,ma2016all,bahari2017nonreciprocal}, which provides the possibility for constructing a large-scale robust photonic circuit on a single chip.

On the other hand, silicon photonics has achieved rapid development in recent years owing to the CMOS-compatibility. Silicon's high refractive index can support highly confined optical mode even with a small bend. However, the backscatter caused by surface or side-wall roughnesses of silicon photonic devices has always been a trouble for the photonic circuits~\cite{ji2017ultra,li2016backscattering}. For example, silicon ring-resonators composed by strip waveguides are important for wavelength division multiplexing (WDM) based on their narrowband resonant modes~\cite{little1997microring,bogaerts2012silicon}. However, if there is a light backscatter arisen from the surface roughnesses, the forward incident propagating mode would be coupled by the backward associated propagating mode, giving rise to the resonance-splitting and the distorted Lorentzian-shaped spectrum~\cite{li2016backscattering}. As a consequence, it is desirable to involve the topological robustness into the silicon photonics to be immune from the structure defect and suppress the light backscatter.

In this letter, we propose and experimentally demonstrate a topological photonic ring-resonator fabricated on a silicon chip. It is designed in valley photonic crystals (VPCs) with a triangular-loop topological edge. The unidirectional transmission and robustness against the structure defects of the resonant modes in the ring are illustrated.  By coupling the topological ring with topological waveguides, wavelength-selective notch and channel-drop filters are illustrated for the potential WDM. This result may extend applications of topological photonics into silicon photonic integrated circuits.

\section{Simulation}\label{sec:02}
Figures 1(a) and (b) display schematic diagrams of the proposed topological ring-resonator designed in 2D-TPCs, which is coupled with a topological waveguide. The arrows in Fig. 1(a) indicate the direction of the energy flow of light in the propagating mode. The basic units of the 2D-TPC include two kinds of honeycomb PC lattices with air-holes in a silicon slab, which are marked by VPC$_1$ and VPC$_2$ in the inset of Fig. 1(b). Both of the PC lattices contain three large air-holes and three small air-holes, but they are carefully arranged with broken inversion symmetry. The corresponding Berry curvature and topological invariant valley Chern indices $C_V$ of the two VPCs are thus calculated to be opposite, i.e., $C_V$\textless 0 for VPC$_1$ and $C_V$\textgreater 0 for VPC$_2$~\cite{he2019silicon}. Detailed bulk band for VPC$_1$ and VPC$_2$ has been calculated and plotted in Fig. 1(c) by using the finite-difference time-domain (FDTD) method with a lattice constant of $a=410$ nm. A wide TE-like bandgap appears when two inequivalent air-holes of the unit cell are adjusted with diameters of $d_1=174$ nm and $d_2=81$ nm, respectively. According to the $bulk-boundary$ correspondence, the interface between VPC$_1$ and VPC$_2$ will exhibit topologically protected robust propagations. In Fig. 1(d), we present the dispersion relationship of the valley-dependent edge states for the staggered interface when VPC$_1$ and VPC$_2$ are constructed together. The edge state is momentum-valley locked with a positive group velocity at K' valley and a negative group velocity at K valley. Only one surface mode can propagate in the forward (K' valley-locked) or backward (K valley-locked) direction according to the valley-dependence. 

We choose this valley pair of 2D-TPC because it opens a TE-like bandgap at K(K') point other than the $\Gamma$ center in the $k$-space. Thus, the topologically protected edge mode could be well confined in the silicon PC slab considering the K and K' valleys are both under the light-cone (the grey area) as displayed in Fig. 1(c). This property is beneficial for lossless transmission and for improving the quality of the topological ring-based devices. Therefore, many efforts have been made to construct the high quality ($Q$) factor ring cavity~\cite{yamaguchi2019gaas,ma2019topological,noh2020experimental,zeng2020electrically,barik2020chiral,mehrabad2020chiral} using the VPCs and successfully contribute to the applications in integrated sources like lasers and emitters~\cite{yamaguchi2019gaas,noh2020experimental,zeng2020electrically,mehrabad2020chiral}. Here, the topological ring-resonator is formed by folding the topological interface into a triangular loop, considering the geometric $C_{3v}$ group symmetry. To facilitate the coupling of the resonant modes in the ring, a topological waveguide at the interface of VPC$_1$ and VPC$_2$ is designed. Note, because of the valley-selected unidirectional propagation of the topological edge mode, the waveguide is connected with the ring directly to satisfy the effective coupling.  

For the light (for example, with a normalized amplitude $E_0$) enters from the input port, as shown in Fig. 1(a), only one topological edge mode transports in the forward direction, as marked by the yellow arrows. The backscatter is suppressed owing to the valley 'locked' feature of the interface. At the connection point between the topological ring cavity and the waveguide, the lattice is carefully arranged as shown in Fig. 1(b). The TPC waveguide is directly connected to the corner of the triangular cavity without a gap. As a result, the incident waveguide can be topologically connected to the waveguide of the cavity with effective coupling. Then the incident light would propagate and split topologically into two routes by following the valley-dependent rule as presented in Fig. 1(a). One part ($\alpha{E_0}$) passes directly through the connection point. The other part ($\beta{E_0}$) turns over a sharp $120^{\circ}$ bending and propagates forward in a clockwise (CW) direction in the ring instead of spreading along the $60^{\circ}$ direction. The light is 'locked' to transport in this specific direction during its propagation in the triangular loop. After a round, under the same valley-dependent rule, part of the light in the ring will merge to the through port and the remaining part continues to participate in a new cycle at the connection point. 

After several cycles, the light in the ring could form stable resonant modes if its wavelengths meet the resonant condition. The resonant modes are confined in the ring-resonator, which results in periodic dips in the transmission spectrum of the output waveguide. The whole process could be described by adopting the traditional superimposition of a multi-beam interference. By assuming a lossless balanced, i.e. a 50:50 beam splitting process at the branch considering the topological property and structural symmetry, the final transmission of the topological ring can be written as 
\begin{equation}
T=\frac{0.5+a_l^2-\sqrt{2}a_l{\rm cos}\delta}{1+0.5a_l^2-\sqrt{2}a_l{\rm cos}\delta}
\end{equation}
where $a_{l}$ and $\delta=6{\pi}n_{eff}L/\lambda$ are the round trip loss coefficient and phase shift, respectively. $n_{eff}$ is the effective guiding index of the propagation mode, $L$ is the side length of the triangular ring and $\lambda$ is the operation wavelength. When the phase shift $\delta$ is a multiple of $2\pi$, the light is on resonance and shows a notch in the transmission spectrum. Note, the resonant mode is topologically protected and shows an opposite routing direction (CW direction) from the ring-resonator composed by a strip waveguide, which has the counter-clockwise (CCW) direction and couples with a bus-waveguide via the evanescent field~\cite{bogaerts2012silicon}.  

\begin{figure}[!t]
	\centering
	\includegraphics[width=\linewidth]{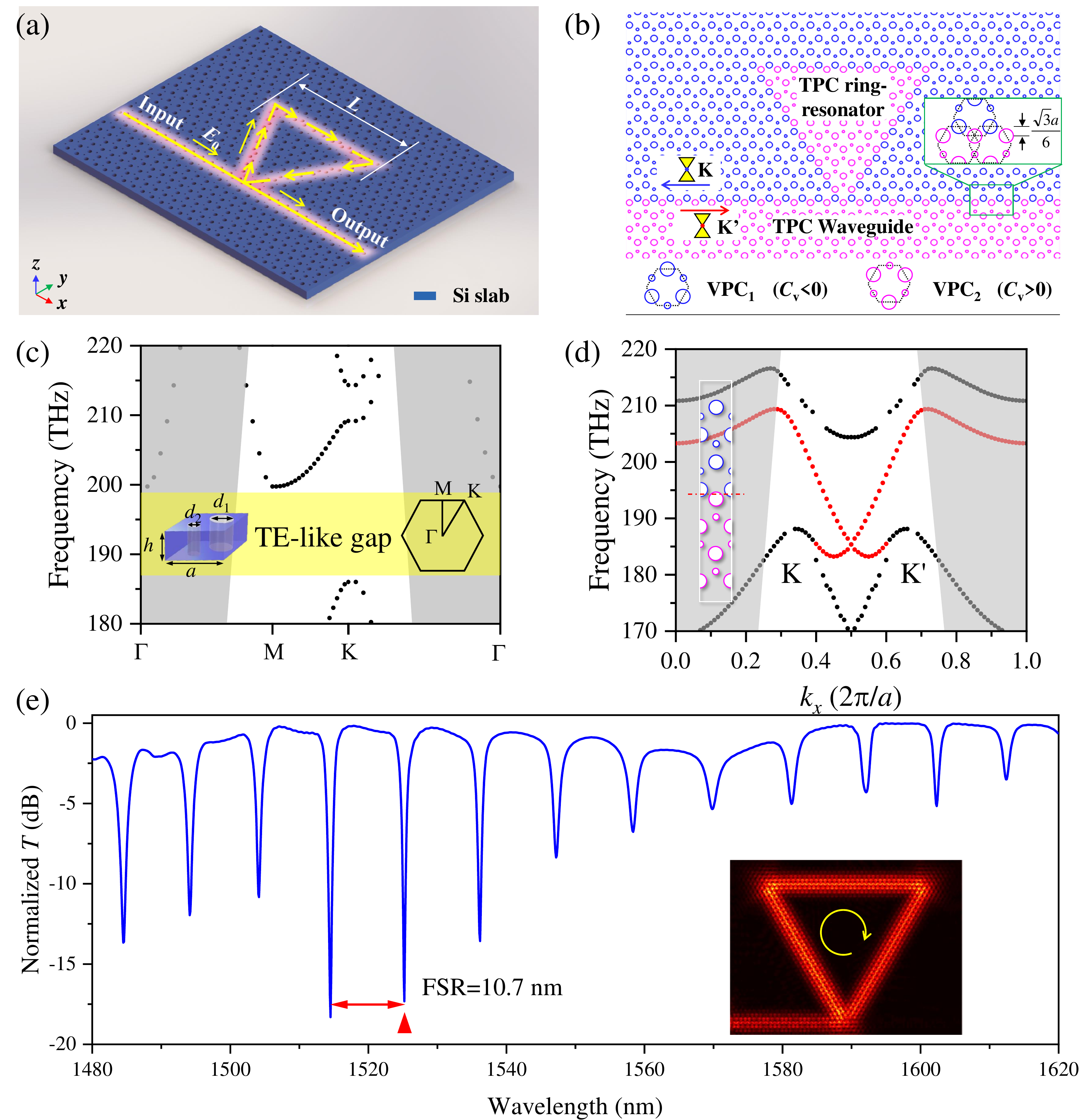}
	\caption{(a) \& (b) Schematic images of a triangular ring-resonator using the VPCs on a silicon slab. (c) Bulk photonic bands for VPC$_1$ and VPC$_2$ as they share the same band structure. (d) Dispersion of the valley-dependent edge states for the staggered interface between VPC$_1$ and VPC$_2$. (e) Simulated normalized transmission spectrum of a VPC triangular ring-resonator. Inset gives the electric field distribution of the ring-resonator at the resonant wavelength of 1525.2 nm.}
	\label{fig:fig1}
\end{figure}

Using the FDTD method, we conduct the simulations to analyse the performance of the topological waveguide-ring structure shown in Fig. 1(a). With the same lattice parameters in the calculation above, a fine-tuned bandgap is desired to match with the telecom band, which is compatible with the laser wavelength in our experiment. In Fig. 1(e), we obtain the simulated normalized transmission spectrum of the proposed structure. It consists of a series of transmission notches at the resonant wavelengths of the triangular ring-resonator. The free spectrum range (FSR) is measured as 10.7 nm in correspondence to the circumference of the cavity ($3L=105a$). By fitting the resonant dips with the standard Lorentzian function, the linewidth of the resonant mode is estimated as 1.26 nm at 1525.2 nm. Therefore, the loaded $Q$ factor is correspondingly calculated as 1,210 with $Q=\lambda/\Delta\lambda$ owing to the large loss induced by the coupling coefficient indicated in Eq. 1. The inset of Fig. 1(c) gives the electric field distribution of the on-resonance state in the waveguide-ring structure at 1525.2 nm, showing a good resonant suppression in the waveguide output.

To verify the light propagation behaviour proposed in Fig. 1(a), which is determined by the topological property of the interface between VPC$_1$ and VPC$_2$, we further simulate the energy flow of the propagation mode. Figure 2(a) gives the poynting vectors of the propagating mode in the whole topological waveguide-ring structure at the resonant wavelength of 1525.2 nm. The incident light is confined in the interface corresponding to the K' valley-locked transmission. A clear way of the energy flow transporting along the CW direction can be observed throughout the whole ring-resonator. Coherent interference occurs at the waveguide-ring connection point (region A, shown in Fig. 2(b)), where no light propagates to the output. The enlarged image in region B (shown in Fig. 2(c)) gives the vanishing field overlapping between the two valley states along the VPC channel, which could greatly resist the defects. In region C, i.e., the corner of the ring triangle, which is zoomed in Fig. 2(d), the flow shows anti-scatter transmission properties. Light transports smoothly along a sharp acute angle with little scatterings due to the topologically protected surroundings of the VPCs~\cite{he2019silicon}. 

To prove the robustness of the topological ring-resonator, an air-hole is intentionally missed at the side-wall of the ring to imitate certain scatters or the presence of the defect. Figure 2(e) presents the same area of region B in case with an air-hole defect. The missing-hole defect has been highlighted with blue dotted curve. Compared with that shown in Fig. 2(c), the flow goes through the defect with no apparent scatterings along the VPC channel. In Fig. 2(f), more air-holes are missed while the directionality of the flow is still maintained well. In addition, in Fig. 2(g), we obtain the simulated resonance dips of the proposed VPC triangular waveguide-ring structures with and without air-hole defects. For the case with one air-hole defect in region B, as shown in Fig. 2(e), there is only a slight shift of the resonance dip without decrease of $Q$ factor. Similar defects in regions A and C are also considered, showing an ignored impact on the resonance dip as well. For the case shown in Fig. 2(f), where three air-holes are missed, the resonance shift become larger. However, when the air-hole defect with the same size of $d_2=81$ nm is dug out from a conventional silicon ring resonator composed by a strip waveguide with a width of 500 nm, an obvious shift of the resonance dip is observed, as shown in Fig. 2(h). Moreover, the backscattering induced by this air-hole defect is enhanced in the ring, resulting in an unwanted splitting of the resonance~\cite{li2016backscattering}. Therefore, we can conclude that the intervalley scattering aroused by the defects is naturally suppressed and this ring-resonator could be designed to against certain perturbations. It is beneficial for future applications to construct backscattering-free photonic circuits.

\begin{figure}[!t]
	\centering
	\includegraphics[width=\linewidth]{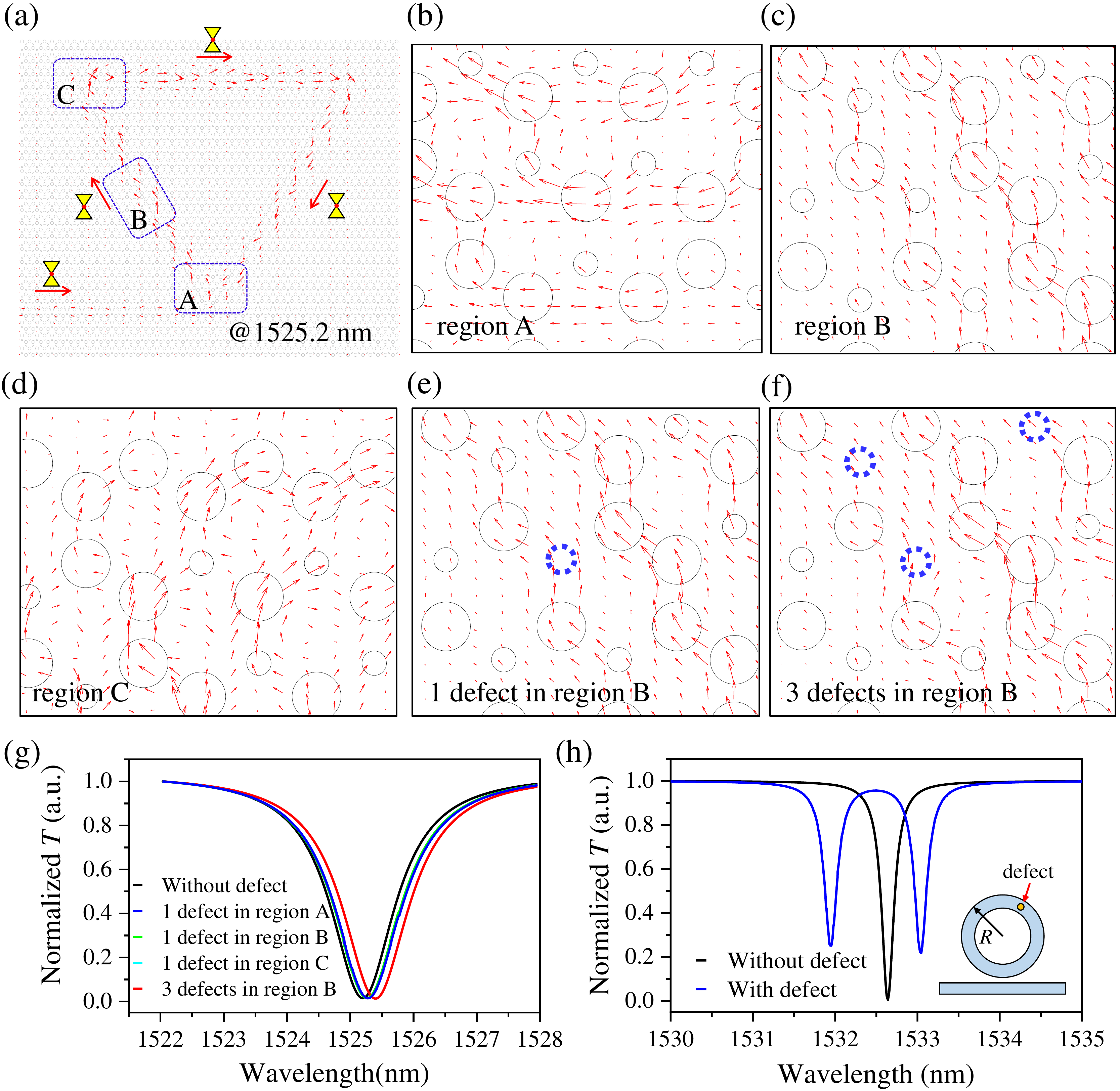}
		\caption{(a) Poynting vector of the resonant mode propagating in the ring-resonator, showing a CW rotation direction. (b) Zoomed image of region A in (a) at the waveguide-ring coupling region. (c) Zoomed image of region B in (a) along the VPC channel. (d) Zoomed image of region C in (a) at the corner of the triangular ring-resonator showing the anti-scatter transmission. (e) \& (f) Zoomed image of region B along the VPC channel with one and three air-hole defects, respectively, showing defect immune in comparison with (c). (g) Simulated resonance dips in the proposed topological ring-resonator with and without air-hole defects in different regions shown in (a). (h) Simulated resonance dips in a conventional ring-resonator (with a radius of 4~$\upmu$m) composed by a strip waveguide with and without an air-hole defect. The inset is a schematic of the ring-resonator with an air-hole defect.}
	\label{fig:fig2}
\end{figure}

\section{Experiment Results}\label{sec:03}
To experimentally verify the proposed topological ring-resonator and demonstrate its function of wavelength-selective filtering, we fabricate it on a silicon-on-insulator with a 220 nm top silicon layer and 2 $\upmu$m buried oxide layer. Electron beam lithography combined with inductively coupled plasma etching are used to accomplish the fabrication. The lattice constant and diameters of the air-holes of the VPCs are designed appropriately with the parameters mentioned above. The side length of the ring is about $L=35a$. To facilitate the light coupling into and out of the topological waveguide-ring, strip waveguides with tapered width are designed at the two ends of the topological waveguide. In addition, the strip waveguides have grating couplers (GCs) at the ends to vertically couple with optical fibers for laser incidence and transmitted power measurement. Figure 3(a) displays an optical microscope image of the fabricated device. To clearly characterize the topological photonic structure, a zoomed scanning electron microscope (SEM) image is shown in Fig. 3(b) as well. The false colors are added to highlight the regions of the inequivalent VPC$_1$ and VPC$_2$.  

A narrowband tunable laser is coupled into one of the grating couplers, which transmits along the tapered waveguide and couple into the topological waveguide and ring. The transmission power of the whole device is monitored from the other grating coupler. By tuning the laser wavelength, the transmission spectrum of the device is obtained, as shown in Fig. 3(c). The transmittance has been normalized by dividing the maximum value obtained from a single TPC waveguide with the same parameters. Periodic transmission notches with the Lorentzian lineshapes are observed. The non-flat transmission background is attributed to the wavelength-dependent coupling efficiency of the grating couplers. The measured FSR is about 10.7 nm with an extinction ratio (ER) close to 20 dB, which is consistent with the simulation result shown in Fig. 1(c). The $Q$ factors are estimated around 1,050, which are close to the simulation results as well. The obtained transmission dips with a narrow bandwidth at the resonant wavelengths indicate their abilities of all-pass notch filter, i.e., light meeting the resonant wavelengths of the ring-resonator will be filtered out and could not transport from the 'Through' port.

In addition, similar to the discussion in Fig. 2, we intentionally miss air-hole defects around the VPC channels in the fabricated devices, as displayed in Fig. 3(d). Corresponding transmission spectra are plotted in Fig. 3(c) as well. Though the missed air-hole defects and the inevitable fabrication errors, the resonant modes of the ring-resonator are not perturbed, showing barely unchanged resonant wavelengths and $Q$ factors. It indicates the topologically protected robust transport and defect immunity of the proposed topological ring-resonator. 

\begin{figure}[!t]
	\centering
	\includegraphics[width=\linewidth]{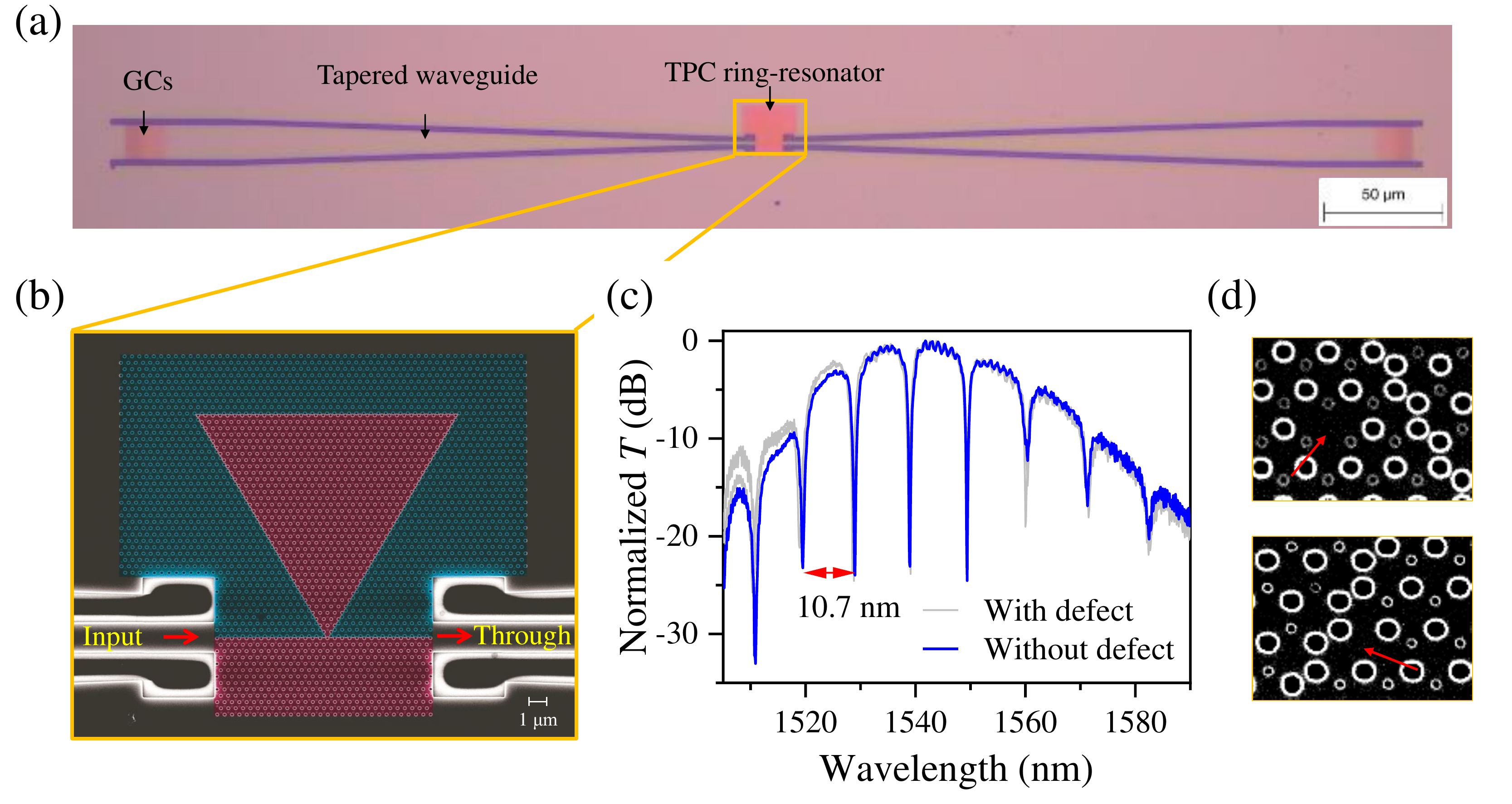}
	\caption{(a) Optical microscope image and (b) zoomed SEM image of the fabricated topological ring-resonator. (c) Measured transmission spectra of the fabricated devices with and without the missed air-hole. (d) Zoomed SEM images around the VPC channels with missed air-holes. }
	\label{fig:fig4}
\end{figure}

The above experiments indicate the possibility of an all-pass notch filter using the proposed topological ring-resonator, as well as its good defect immunity property. Next, we illustrate that this structure could also be employed to achieve a channel-drop filter. As shown in Fig. 4(a), a second topological channel waveguide is added at another corner of the triangular ring. To satisfy the topologically protected guiding mode at the VPC interfaces and effectively couple with the resonant mode in the ring-resonator, this channel waveguide has an angle of 60 $^\circ$ with respective to the two sides of the triangular ring and connects with the corner of the ring-resonator directly. For the on-resonance light incident from the input topological channel waveguide, it couples into the ring-resonator and circulates to constructively build up in energy. The spectrum from the direct output end of the input waveguide ('Through' port) will exhibit transmission dips owing to the destructive interference and function as the notch filter, as discussed above. The resonant mode circulating in the ring will couple into the second topological channel waveguide, which is similar to the coupling process between the ring and the input topological waveguide discussed in Fig. 1. Differently, coherent transmission peaks will come from the output of the second channel waveguide without the interference with the primary input wave. Consequently, the output end of the second topological channel waveguide could be considered as a 'Drop' port for functioning as a channel-drop filter to select the optical signal at specific wavelengths, which is similar to the definition in the ring-resonator composed by strip waveguide~\cite{little1997microring}.  

With the parameters of the VPCs employed in Fig. 3, we fabricate the channel-drop filter, as shown in Figs. 4(a) and (b). The transmission spectra from the 'Through' port and 'Drop' port are both measured and plotted in Fig. 4(c). Similar as the results shown in Fig. 3(c), the 'Through' port of the filter presents periodic transmission dips at the resonant wavelengths. The FSR remains consistent since the ring length is unchanged. However, the linewidths of the resonant dips are broadened and the ERs are reduced. This could be attributed to the addition of the second channel waveguide that reduces the confinement of the resonant mode in the ring via the extra waveguide-coupling loss. Correspondingly,  the spectrum from the 'Drop' port shows transmission peaks at the resonant wavelengths with the ER of more than 10 dB, i.e., light at specific wavelengths are selected via the topological ring-resonator. 

\begin{figure}[!t]
	\centering
	\includegraphics[width=\linewidth]{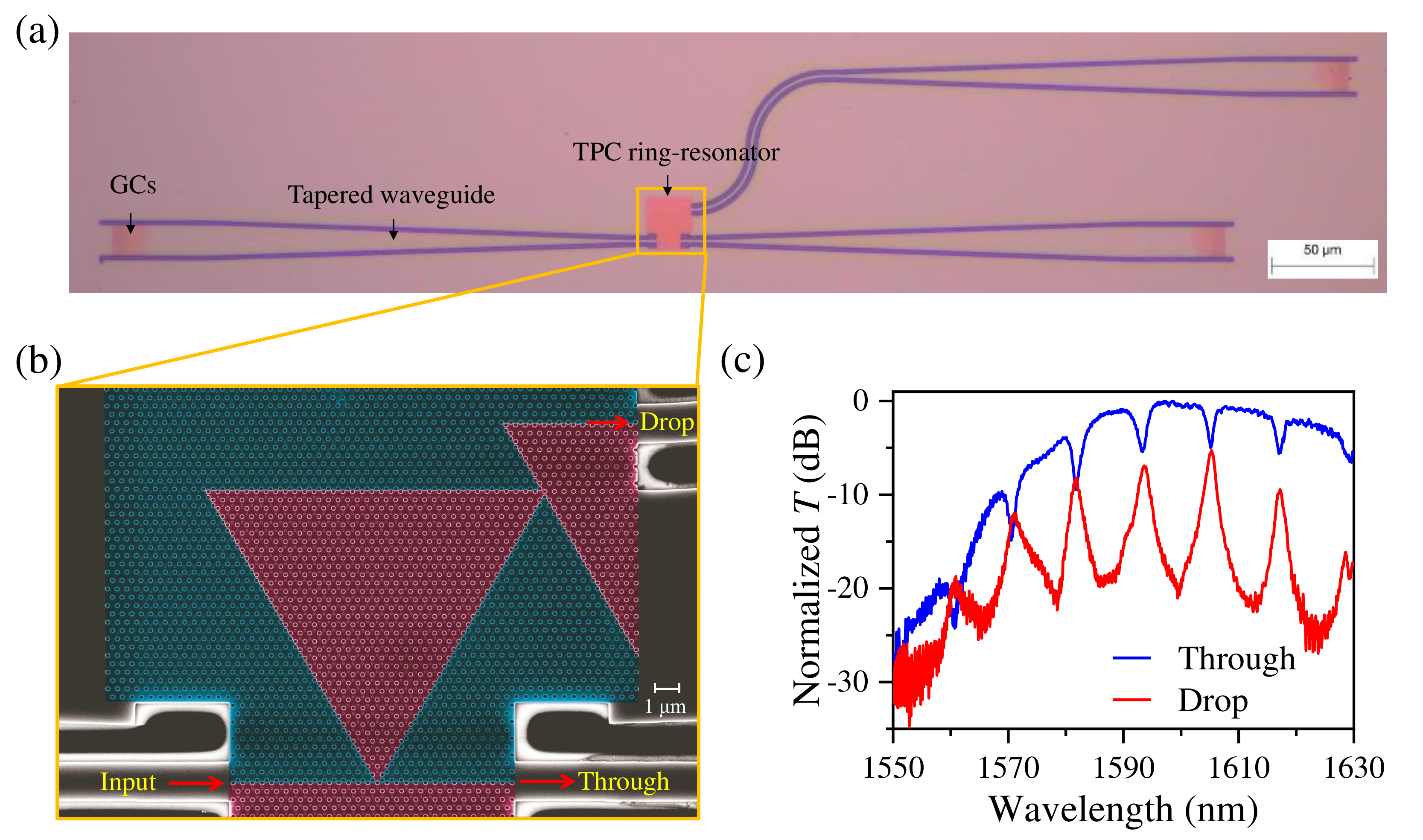}
	\caption{(a) Optical microscope image and (b) zoomed SEM image of a channel-drop filter. (c) Measured transmission spectra from the 'Through' port and 'Drop' port, respectively.}
	\label{fig:fig3}
\end{figure}

\section{Conclusion}\label{sec:04}
In conclusion, we have successfully demonstrated a topological ring-resonator fabricated on a silicon chip. It is constructed by folding a topological edge between two VPCs into a triangular loop, which supports resonant modes with unidirectional transmission and robustness against structure defects. Considering the ring-resonator is one of the most important on-chip photonic elements, the realization of the topological ring-resonator indicates the topological photonics could provide a new platform immune from material and structure defects for developing advanced photonic integrated circuits. As an example, we present notch and channel-drop filters by coupling the topological ring-resonator with topological waveguides. Although the realization of nano-scale air-holes in VPCs is currently a challenge as it requires high-resolution nano-fabrication technique, the topological protection promises the proposed filters to be applied in more complex integrated devices with higher performance such as cascaded high-order filters and low crosstalk WDM devices.


%

%

~\\
\section*{Fundings}\label{sec:05}
the Key Research and Development Program 2017YFA0303800, in part by the National Natural Science Foundation 91950119, 11634010, 61775183, 61905196, in part by the Key Research and Development Program in Shaanxi Province of China 2020JZ-10, in part by the Fundamental Research Funds for the Central Universities 310201911cx032, 3102019JC008, and in part by the seed Foundation of Innovation and Creation for Graduate Students in Northwestern Polytechnical University CX2020208.

~\\
\section*{Acknowledgment}\label{sec:06}
The authors would thank the Analytical \& Testing Center of NPU for the assistances of device fabrication.



\end{document}